\documentclass[journal]{IEEEtran}
\usepackage{xcolor}
\usepackage[linesnumbered,ruled,vlined]{algorithm2e}
\usepackage{algorithmic}

\SetCommentSty{mycommfont}
\SetKwInput{KwInput}{Input}
\SetKwInput{KwInitialize}{Initialize}
\SetKwInput{KwOutput}{Output} 
\usepackage{xcolor,soul,framed} 
\usepackage{xspace}
\colorlet{shadecolor}{yellow}
\usepackage[pdftex]{graphicx}
\graphicspath{{../pdf/}{../jpeg/}}
\DeclareGraphicsExtensions{.pdf,.jpeg,.png}
\usepackage[cmex10]{amsmath}
\usepackage{amssymb}
\usepackage{array}
\usepackage{mdwmath}
\usepackage{mdwtab}
\usepackage{eqparbox}
\usepackage{url}
\usepackage{cite}
\usepackage{multirow}
\begin{document}
\bstctlcite{IEEEexample:BSTcontrol}
    \title{A Drone-Aided Blockchain-Based Smart Vehicular Network}
  \author{Muhammad Asaad~Cheema,
      Muhammad Karam~Shehzad,\\
      Hassaan Khaliq~Qureshi,~\IEEEmembership{Senior Member,~IEEE,}
      Syed Ali~Hassan,~\IEEEmembership{Senior Member,~IEEE,}
     and~Haejoon~Jung,~\IEEEmembership{Member,~IEEE}

  \thanks{M. A. Cheema, M. K. Shehzad, H. K. Qureshi, and S. A. Hassan are with the School of Electrical Engineering \& Computer Science (SEECS), National University of Sciences \& Technology (NUST), Islamabad, Pakistan. emails: (mshehzad.msee17seecs, mcheema.msee18seecs, hassaan.khaliq, ali.hassan)@seecs.edu.pk.}
  \thanks{H. Jung is with the Department of Information and Telecommunication
Engineering, Incheon National University, Incheon 22012, South Korea. e-mail: (haejoonjung@inu.ac.kr).}
}


\maketitle

\begin{abstract}
The staggering growth of the number of vehicles worldwide has become a critical challenge resulting in tragic incidents, environment pollution, congestion, etc. Therefore, one of the promising approaches is to design a smart vehicular system as it is beneficial to drive safely. Present vehicular system lacks data reliability, security, and easy deployment. Motivated by these issues, this paper addresses a drone-enabled intelligent vehicular system, which is secure, easy to deploy and reliable in quality. Nevertheless, an increase in the number of operating drones in the communication networks makes them more vulnerable towards the cyber-attacks, which can completely sabotage the communication infrastructure. To tackle these problems, we propose a blockchain-based registration and authentication system for the entities such as drones, smart vehicles (SVs) and roadside units (RSUs). This paper is mainly focused on the blockchain-based secure system design and  the optimal placement of drones to improve the spectral efficiency of the overall network. In particular, we investigate the association of RSUs with the drones by considering multiple communication-related factors such as available bandwidth, maximum number of links a drone can support, and backhaul limitations. We show that the proposed model can easily be overlaid on the current vehicular network reaping benefits of secure and reliable communications.   
\end{abstract}

\begin{IEEEkeywords}
Backhaul network, blockchain, drones, intelligent transportation, roadside units (RSUs), UAVs, unsupervised learning.
\end{IEEEkeywords}

%
\IEEEpeerreviewmaketitle


\section{Introduction}
\IEEEPARstart{T}{remendous} increase of global Internet traffic has drawn the attention of researchers both from academia and industry. This exponential increase of Internet traffic demands high data rates with low latency, which can be viewed as a challenging problem in smart vehicular networks. Considering the evolution of fifth-generation (5G) and beyond-5G (B5G), and catering the coverage issues to provide reliable communication, unmanned aerial vehicles (UAVs), drones, or unmanned balloons (UBs) have emerged as a promising solution owing to their on-demand deployment, relocation, autonomy, good line-of-sight (LoS) access to unreachable areas, and low-cost characteristics \cite{drone_aided,uav_2,uav_3, a_1, karam_kmeans}.

Due to such advantages, one of the key applications of this technology is in the domain of smart traffic system \cite{vehicular_drone}. Statistics indicate that within a short span of time, the staggering growth of vehicles worldwide resulted in environmental pollution, congestion, tragic incidents and fatal accidents, which had directly affected the human societies. Therefore, the concept of vehicular communication has been introduced to lessen such concerns \cite{secure_vehicular}. However, the current vehicular communication technology lacks reliable, secure, and energy-efficient communications to incorporate intelligent transportation system. Therefore, there is a gap of introducing novel techniques for secure, reliable and energy-efficient communication technologies for drone-enabled smart vehicular networks (DESVNs). 

\begin{figure*} 
\begin{center}
  \includegraphics[width=\textwidth,height=7cm]{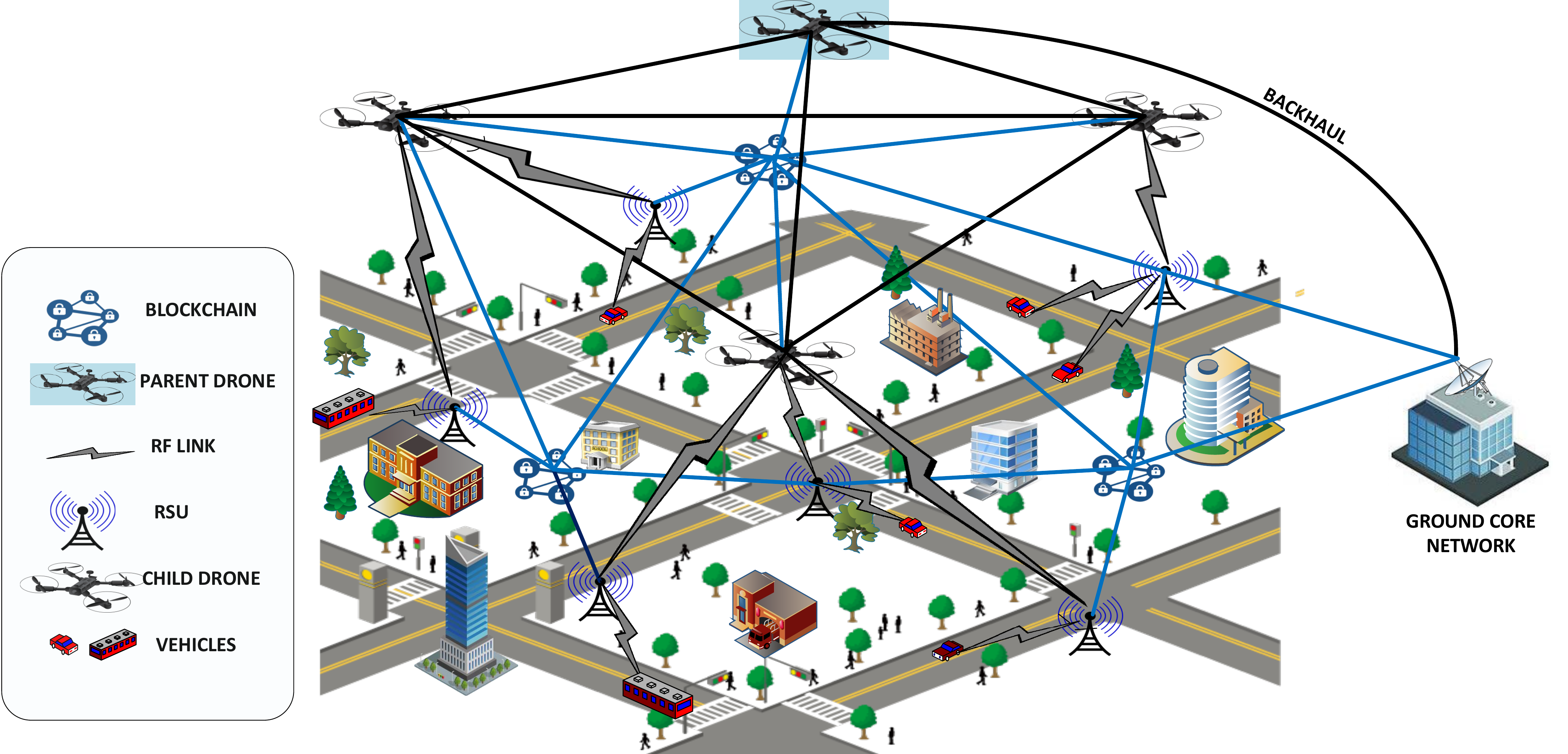}
  \caption{{Smart vehicular network scenarios.}}\label{systemmodel}      
\end{center}
\end{figure*}

Keeping in view the on-demand and cost effectiveness of the network, drone-enabled wireless technology can play a vital role. However, lack of the secure connection and mutual trust between drones makes the drone-enabled wireless networks vulnerable to different attacks specially, when the drones are working autonomously and able to make decisions on their own \cite{a_2}. In this context, security becomes a critical issue in the drone-enabled wireless networks. In fact, there are different types of attacks including unauthorized access to the system, allocating resources to the malicious users, and disrupting the availability of resources. Not only do these attacks sabotage the communication for target infrastructure, but they can also become a catalyst for bigger attacks \cite{a_3}, \cite{a_4}. Therefore, it is important to come up with a secure infrastructure for drone communication, where drones can communicate with mutual trust and security. Further, reliability and energy efficiency of such a network should be optimized at the same time.

To provide mutual trust and security to the systems involving the deployment of drone technologies, various approaches are proposed in the existing studies. For example,  in \cite{L1}, the authors suggest the use of artificial intelligence to predict behavior of the UAVs. A recurrent neural network is used for the authentication of the signals by analyzing the previous security states of a UAV to protect the system. Similarly, an intrusion detection scheme is proposed in \cite{L2}, in which the authors use the concept of support vector machines to categorize the UAVs as normal or malicious to provide the security to the network. Moreover, other intrusion detection techniques in drone environment have also been proposed, as in \cite{L3}. Most of these techniques employ a centralized way of protecting the system environment involving drones. 

On the other hand, with the advent of blockchain technology, a great interest has been witnessed recently because of its integrity and distributed nature, and leverage to provide reliable communication \cite{reliable}. Therefore, researchers adopt blockchain technology and further improve the system architecture by resolving various blockchain limitations \cite{light}. Blockchain is a chain of blocks involving cryptographic algorithms where the block header contains the hash of the previous block, which adds an extra layer of security and integrity \cite{a_5}. Originally, blockchain was designed for financial purposes to exchange digital currency without the involvement of third party. However, many other applications are emerging such as healthcare, supply chain, and logistics that use blockchain, because of its decentralized and audit-able nature, where every participant in the network can add reliable data to the blockchain \cite{a_6,a_7,a_8}.

Due to the distributed nature and cryptographic abilities of blockchain, it can be a perfect ingredient for security, which provides the mutual trust between different entities of the system and handles the authentication of all the entities in the system. The authors in in \cite{L4} propose an intelligent approach for UAVs by using the concept of blockchain. In this technique, the authors use the concept of encryption and decryption exploiting the public/private scheme to protect the signals from controller to drone. Similarly, the authors in \cite{L5} propose the usage of drones in a secure way as on-demand nodes for inter-service operability between multiple vendors by using the concept of blockchain. A blockchain-based network flying platforms (NFP) are investigated in \cite{L6}, where the authors study how to use blockchain to increase the resilience of NFP. 

Different from the existing studies, in this paper, a blockchain-based secure intelligent transportation system is proposed for the drone-enabled wireless technology. We integrate the concept of blockchain between different entities of the network to ensure mutual trust. Also, we propose a blockchain-based registration and authentication mechanism, which not only builds the mutual trust between the different entities of the system but also avoids the single point of failure and denial of service due to the distributed nature of the blockchain. The mechanism involved here is responsible for the registration of the entities such as smart vehicles (SVs), roadside units (RSUs) and drones on the blockchain and later getting the registration information from the blockchain to authenticate the available RSUs and drones as well the requested SVs. From the prospective of drone communication, energy-efficient communication of drones with the ground RSUs is addressed. In particular, efficient positioning of drones is discussed and then association of RSUs with the drones is evaluated with the objective of maximizing the sum-rate of the overall network. In addition, optimum power allocation of drones is taken into account by considering interference as noise \cite{opt_1}, so that the drones can service for a longer period of time \cite{hover_time, uav_control}.

The rest of the paper is organized as follows. In Section \ref{architecture}, we present the architecture of DESVN. Section \ref{proposed} discusses our proposed model for DESVN. In the same section, we describe security of proposed network, positioning of drones, and path loss model related to drone-based communication. Section \ref{objectivenet} focuses on the objective formulation of drone-based communication. Also, the association of RSUs with drones is discussed in the same section. Performance evaluation of the proposed model is presented and explained in Section \ref{performance}. Finally, Section \ref{conclusion} concludes the paper and addresses future extensions.    
\section{DESVN's Architecture}\label{architecture}


\par Fig.~\ref{systemmodel} shows a graphical illustration of the DESVN for a typical urban area. Mainly, there are four network entities; SVs, RSUs, ground core-network, and drones. There are two classes of the drones. Based on their height, the first class of drones is the child-class, which are hovering at an altitude of $h_D$ from ground level. Further, these drones are capable of communicating (e.g., exchanging control information) with each other through a free space optical (FSO) communication link\footnote{FSO-based links are assumed ideal in our work, which means no losses.}. Moreover, child-drones are playing the role of access points for the RSUs using radio frequency (RF) signaling. The other class of drones is classified as the parent-drone class, which are flying at an altitude higher than the child-drones, so that a perfect LoS communication with the ground core-network and child-drones is possible. Without loss of generality, there is only one parent-drone considered in this study to simplify the network\footnote{However, multiple parents can be considered as well and we leave this as future work.}. The communication between the child-drones and parent-drone is assumed to use another FSO link. In addition, the parent-drone is communicating with the ground core-network via an FSO link as well. The idea of using such an architecture is to reduce the complexity of the overall network. For instance, with the introduction of a parent-drone, multiple backhaul links are eliminated. Further, the parent-drone is beneficial to make a perfect LoS communication link with the ground core-network. On the other hand, the child-drones are responsible for delivering the collected information of their region to the parent-drone, which indeed simplifies the task of the parent-drone. Also, the child-drones are not required to communicate with the core-network, which reduces the complexity of their communications and results in eliminating the requirements of the backhaul link. We believe that such an architecture can play a vital role in the smart vehicular networks, where the on-spot delivery of data is of pivotal importance.

RSUs are directly communicating with the SVs on the road, and they are forwarding the gathered information to the respective child-drone operating in their regions. We assume that the blockchain technology is integrated in between RSUs and drones, maintained and managed by the ground core-network. The concept of the blockchain is used to register these two entities alongside with SVs, so that it would be able to verify and authenticate these entities. This is done to avoid the malicious entities to get into the system and to build mutual trust between these entities, so that they can share data with trust and ease. Also, in the system model, it is assumed that SVs already contain the latest and verified copy of the blockchain. An Ethereum platform is used for interaction with the blockchain and for the deployment of the smart contract. In addition, it is also used for the registration of the network entities. However, in this work, only security-related concerns of the RSUs and SVs are taken into account. Therefore, the communication paths between the RSUs and SVs are assumed to be ideal, which means no losses. Without loss of generality, we consider a three-dimensional (3D) Cartesian coordinate system, in which the locations of RSUs are denoted by ($x_i, y_i$) with zero altitude, while the locations of the child-drones as ($x_{Dj}, y_{Dj}, h_{Dj}$), where $i\in\{1, 2, ..., U\}$ and $j\in\{1, 2, ..., V\}$. Furthermore, the parent-drone is represented by a singleton set $P_0$. With such an architecture, we will introduce our proposed model for the DESVN in the following section.

\section{Proposed Model for smart traffic environment}\label{proposed}
\begin{figure} 
\begin{center}
  \includegraphics[width=8.5cm]{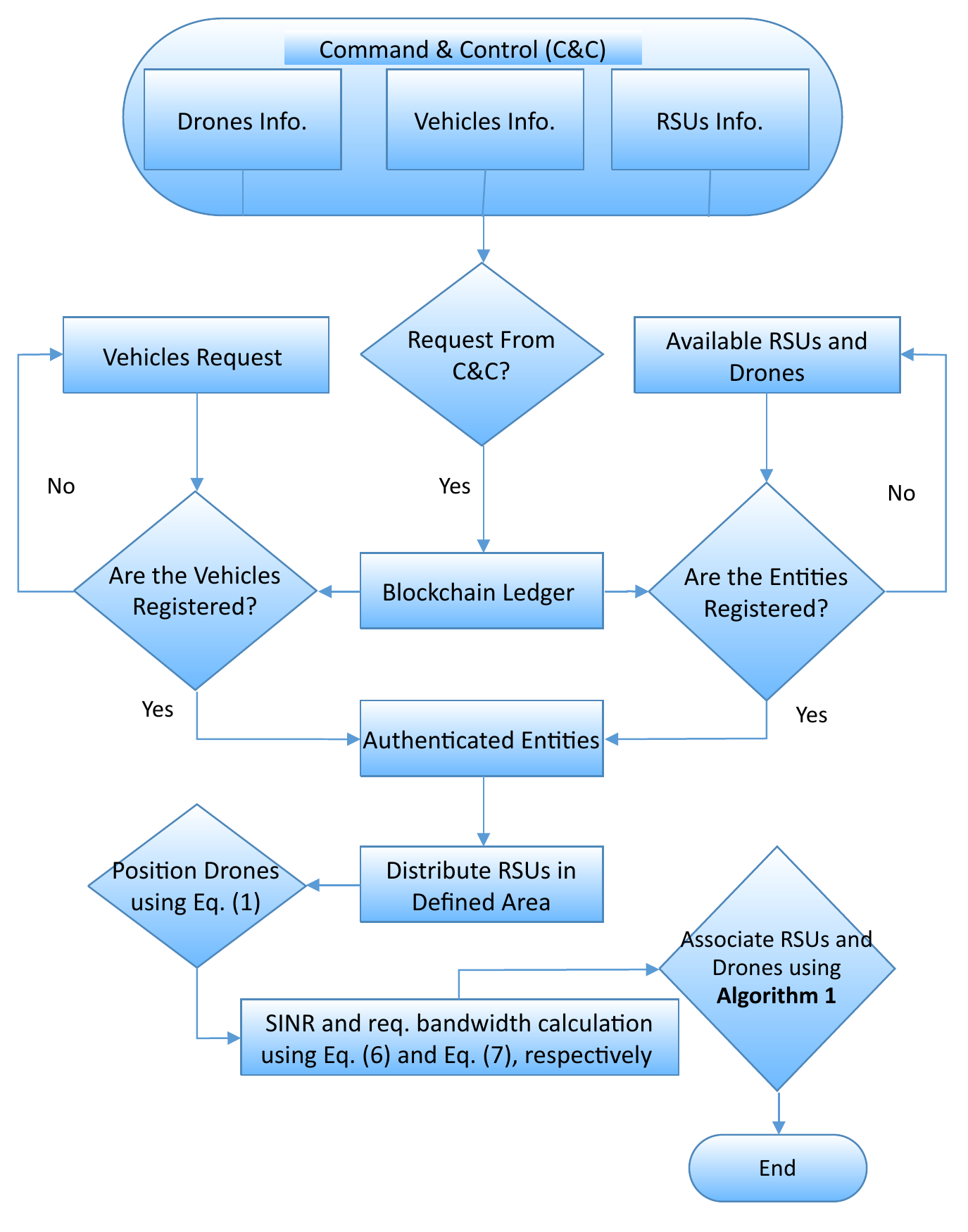}
  \caption{The flow graph of the proposed model.}\label{flowchart}
\end{center}
\end{figure}

Fig.~\ref{flowchart} illustrates a flow chart of our proposed work, i.e., a blockchain-based registration and authentication system of the drones, SVs, and RSUs. The figure shows that the information for the registration process should come from the command and control (C\&C). If the information related to entities is arriving from C\&C,  it is considered authentic and is mapped against a unique ID. On the other hand, if the registration request is not from C\&C, then the information will be discarded, and the registration of the entity is rejected. Further, the flow chart depicts the authentication mechanism, which involves the extraction of information from the blockchain to authenticate the entities in the network. The entities provide their information in order to be authenticated. If the entity is registered by the C\&C at the first place and the information they are providing is correct (i.e., the provided information matches with the information already stored on the blockchain), then the entities can authenticate themselves. Only those available entities, which are registered on the blockchain and  authenticated by providing credentials, are allowed to take part in further steps. Therefore, the authenticated RSUs are distributed in the given area. Later on, only the authenticated drones are positioned, and the required parameters are calculated, which are then passed to Algorithm\,\ref{algo} for the association of RSUs with drones. In the following subsections, we first describe the authentication and registration process, and then we introduce the positioning of drones and path loss model in detail. 
\begin{figure} 
\centering
  \includegraphics[width=8.8cm]{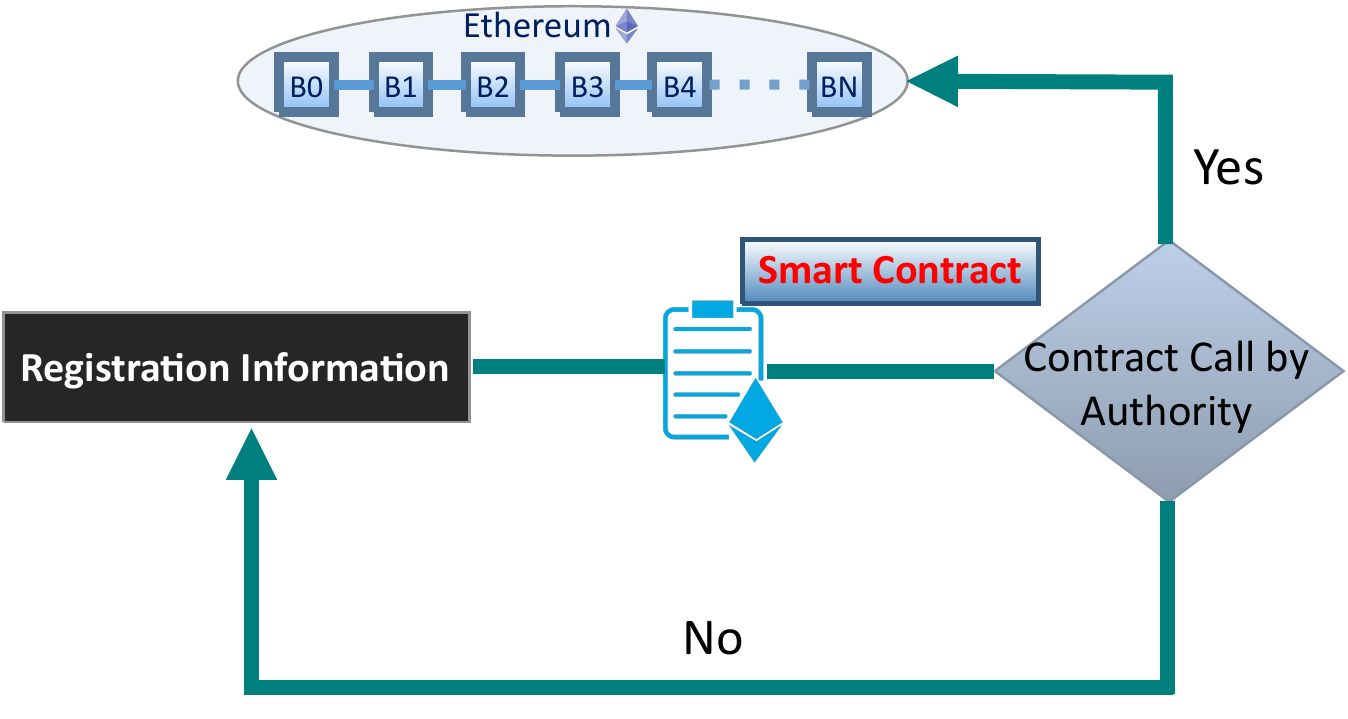}
  \caption{{A snapshot of registration mechanism.}}
  \label{Blockchainfig}
\end{figure}

\subsection{Blockchain}
A blockchain network is deployed between the RSUs and the drones. The blockchain network is maintained and managed by the blockchain center, which in this case is considered to be the part of ground core network. In this system, every entity, which is part of the blockchain network, contains a 20 Byte address, through which it can store/retrieve its information to/from the blockchain. The blockchain network is used to share the authenticated information of the registered entities once they get registered by using the following mechanism as shown in Fig.~\ref{Blockchainfig}. For the registration purpose, multiple functions are designed, collecting the information related to different entities as an input and mapping it against the blockchain address associated with each entity. Once the information gets on to the blockchain, each entity, which is connected to the blockchain, can access this information and can verify the source of the information.

\subsection{Registration and Authentication Process of RSUs, Drones and SVs}
As we are using the concept of the blockchain for our registration and authentication process, a smart contract is needed to interact with the blockchain in a sophisticated manner. For this reason, a smart contract is designed with several functions for the registration of SVs, drones, and RSUs, which involve storing data on the blockchain. It also includes functions for the authentication, which is required to get data back from the blockchain.

\subsubsection{Registration Process}
Registration process use three main functions of the smart contract, which correspond to the registrations of drones, RSUs, and SVs, respectively. The reason of designing three different functions separately in smart contract is that each entity contains different set of information associated to it. As defined in the system model, each entity is considered to be the part of the blockchain network, where each entity has the 20 Bytes unique address associated with it on the blockchain network.

The function used for the registration of the drones takes the ID, which is required for the identification of the parent-drone or the child-drones, and allows flying area code of the drone as an input, which is mapped against the blockchain address associated to the drone. Similarly, for the RSUs the function only takes the deployed area of the RSU as an input and stores this information against the 20 Byte address associated to RSU. In the case of the SVs, the smart contract registers them just by taking blockchain address associated with each vehicle and makes SVs eligible for the use of the resources.

Moreover, the smart contract is designed in such an intelligent way that only the C\&C can register these entities by making transactions on the blockchain to update state variables associated to each unique address. In case that the transaction is not made by the C\&C, the transaction is turned down without registering any entity on the blockchain. In addition, the entities connected to the blockchain update their blockchain storage and permit information, which comes from the trusted authority. In order to keep the required gas for each transaction in a specific limit, we keep the information associated to each registration request  within specific length defined by C\&C. Hence, the ID attached to each drone can have maximum 5-character string, and the flying area code can have string length up to 4-characters. Similarly, RSUs can have 4-character string for their deployed area information. As the blockchain has the distributed nature, each entity can get the registration information connected to the blockchain.

\subsubsection{Authentication Process}
Authentication of drones, RSUs, and SVs is shown in the second part of the flow diagram. The SVs request for the authentication by providing the unique addresses associated to them. If the addresses is in the list of the registered vehicles addresses, which is verified by the blockchain, then they are authenticated and allowed to use the resources. On the other hand, we have the available RSUs and drones, before utilization of these entities to provide the resources to the users, authentication needs to be done. Available RSUs and drones provide the unique 20 Bytes addresses associated to these entities in a similar manner as provided by the smart vehicle. These entities get authenticated, if the addresses associated to them gets a match with provided registered entities from the blockchain by using the call function designed in the smart contract for these two entities. In contrast, if the provided address does not match, then the blockchain algorithm rejects the request to protect the system from unauthorized and malicious entities.

{\subsection{Distribution of RSUs and Positioning of Drones}}\label{kmeans}
We assume that the RSUs are distributed according to $\textit{Matern type-I}$ hard-core process \cite{matern_pro} with a density of $\delta$ per meter square, and the minimum distance between the two RSUs is $\zeta_{RSU}$ in meters. The positioning of child-drones is obtained using the $K$-means clustering algorithm \cite{karam_kmeans}, \cite{k_means}. The $K$-means partitions the distributed RSUs into $K$ clusters, where each RSU belongs to a cluster with the nearest mean. Without loss of generality, suppose $\mathcal{R}=\{\mathbf{R_1, R_2, R_3, ..., R_U}\}$ be the set of RSUs, where each $\mathbf{R}_u$, $u\in \{1, 2, 3, ..., U\}$ is a two dimensional vector, which corresponds to the $(x, y)$ location of the $u^{th}$ RSU. Therefore, the unsupervised $K$-means algorithm partitions the $U$ RSUs into $K$ clusters such that $\mathcal{C}=\{C_1, C_2, C_3, ..., C_K\}$ becomes the set of clusters, where $K<U$. Mathematically, it can be expressed as
\begin{equation}
\underset{\mathcal{C}}{\operatorname{\textbf{arg}\,min}}\sum_{l=1}^{K}\sum_{\mathcal{R}\in {{C}}_l }{\mid\mid{\mathcal{R}-{\bf{G}}_{l}\mid\mid^2}}
 \label{eq:8},
\end{equation}
where $l=\{1, 2, 3, ..., K\}$. Also, ${\bf{G}}_l$ represents a two dimensional vector that corresponds to the mean of set of the RSUs deployed in cluster $C_l$, and thus it is the $(x, y)$ coordinates of the cluster. Therefore, \eqref{eq:8} iteratively finds the location of the centroids. Finally, the child-drones are placed on the obtained centroids at the altitude of $h_{D}$. Thus, each child-drone has a number of candidates (RSUs) in its own cluster.

\subsection{Path Loss Model}
Considering the distribution of the child-drones and the RSUs, the horizontal distance between the $i^{th}$ RSU and the $j^{th}$ child-drone is represented as
\begin{equation}\label{eq:1}
{s}_{{i},{j}}= \sqrt{{(x_i-x_{Dj})}^2+{(y_i-y_{Dj})}^2}\: \:.
\end{equation}
The probability of LoS between the $i^{th}$ RSU and the $j^{th}$ child-drone is a fundamental factor in the path-loss calculation. Following \cite{k_6} and \cite{k_7}, this  LoS probability is given by
\begin{equation}\label{}
{\varrho}^{L}_{i,j} = \frac{1}{1+\alpha\cdot\exp\big\{-\beta(\theta-\alpha)\big\}}\:\:,
\end{equation}
where $\alpha$ and $\beta$ are environment constants, and $\theta$ is the angle (in degrees) between the child-drone and the RSU. Also, the probability of non-LoS (NLoS) is ${\varrho}^{N}_{i,j}=1-{\varrho}^{L}_{i,j}$. Further, the angle of elevation is calculated as $\theta=\arctan\Big(\frac{h_{Dj}}{s_{i,j}}\Big)$.  

The air-to-ground (ATG) path loss model \cite{k_6}, \cite{k_7} along with fading, $\psi$, for the communication between the $i^{th}$ RSU and the $j^{th}$ child-drone is presented as
\begin{equation}\label{}
\varpi_{{i},{j}} = {F_{0} + {\varrho}^{L}_{i,j}\cdot\varepsilon^{L}+{\varrho}^{N}_{i,j}\cdot\varepsilon^{N} - {\psi}} ,
\end{equation}
where $F_{0}=20 \log_{10}\Big(\frac{4\pi\cdot d_{{i},{j}}}{\lambda_{c}}\Big)$ is the free-space path loss (FSPL). Also, $\varepsilon^{L}$ and $\varepsilon^{N}$ are the additional losses for the LoS and NLoS links, respectively. In addition, $d_{i,j}=\sqrt{{h}^2_{Dj}+s^{2}_{i,j}}$. Further, $\psi_{i,j}$ is expressed as
\begin{equation}\label{}
\psi_{i,j} \text{[dB]}={\varrho}^{L}_{i,j}\cdot\xi_{0}+{\varrho}^{N}_{i,j}\cdot\xi_{1},
\end{equation}
where the envelopes of $\xi_{0}$ and $\xi_{1}$ follow \textit{Nakagami}
distributions. In other words, $|\xi_{\iota}| \sim Nakagami(m)$, where $\iota=\{0, 1\}$, and $m$ is the shape parameter, which takes the value 1 for Rayleigh fading and 4 for Rician fading.

Lastly, the signal-to-interference-plus-noise ratio $(\text{SINR})$ between the $i^{th}$ RSU and the $j^{th}$ child-drone is determined as
\begin{equation} \label{sinr}
{\text{SINR}}_{i,j}=\frac{\vartheta_{i,j}}{\sigma_{n}^{2}+I},
\end{equation}
where $\vartheta_{i,j}$ is the received power at the $i^{th}$ RSU from the $j^{th}$ child-drone, and $\sigma_{{n}}^{2}$ is the noise power. Also, $I$ represents the sum of the interference from remaining $(V-1)$ child-drones.

In addition, the required bandwidth of $i^{th}$ RSU from $j^{th}$ child-drone is calculated as
\begin{equation}\label{eqbw} 
{w}_{i,j}=\frac{r_{i,j}}{\log_{2}{(1+\text{SINR}_{i,j})}},
\end{equation}
where  $r_{i,j}$ is the requested data rate of $i^{th}$ RSU from $j^{th}$ child-drone.

\section{Objective Formulation and Association of RSUs with Drones}\label{objectivenet}
In this section, we first formulate the optimization problem for DESVN. Then, the association of RSUs with drones is explained in detail.

\subsection{Objective Formulation}
Since the RSUs are distributed according to the $\textit{Matern type-I}$ hard-core process, the child-drones can be positioned based on \eqref{eq:8}. We consider a downlink transmission scenario, in which RSUs are downloading the data from ground core-network via the child-drones. The first communication path between the child-drones and the RSUs is limited by a number of constraints including the maximum bandwidth $W_{j}$ available at a child-drone to distribute to the RSUs, the maximum number of requests $\tau_{j}$ that child-drone can serve, the maximum power $P_{max}$ with which a child-drone  transmit while maximizing energy efficiency (defined later in this section), the minimum $\text{SINR}_{\text{min}}$ criteria to satisfy quality-of-service (QoS) requirement, finally, a RSU will only be served by a one child-drone. The second communication path, which is between the parent-drone and the ground core network, is limited by a data rate limit, defined as backhaul data rate $B_{R}$. 
Keeping in view the communication-related constraints discussed above, the objective is to find the best possible association of RSUs with the child-drones such that sum-rate $S_{R}$ of the overall network can be maximized. Mathematically, the problem can be formulated as
\begin{subequations}\label{objective}
\begin{align}
 \begin{split}
 \max_{\{c_{i,j}\}} {\sum_{i=1}^{{U}}}\sum_{j=1}^{{V}} {r}_{i,j}\cdot c_{i,j}
\end{split}\\
\text{s.t.} \hspace{10pt}
\begin{split} \label{bwconstraint}
 & {\sum_{i=1}^{{U}}} {{w}_{i,j}}.c_{i,j} \leq {W}_{j}, \qquad\qquad\qquad\quad\quad  \forall_{j}
\end{split}\\
\begin{split}\label{linkconstraint}
{\sum_{i=1}^{{U}}} c_{i,j} \leq \tau_j, \qquad\qquad\qquad\quad\qquad\quad\:\: \forall_{j}
\end{split}\\
\begin{split} \label{powerconstraint}
p_{{i,j}} \leq {P}_{j}^{max}, \qquad\qquad\qquad\quad\quad\quad \quad\: \forall_{j}
\end{split}\\
\begin{split} \label{Intconstraint}
     c_{i, j}{\cdot \zeta_{i,j} \cdot{p}_{i,j}} \leq I \:, 
\end{split}\\
\begin{split} \label{SINRconstraint}
\text{SINR}_{i,j} \cdot c_{i,j} \geq  \text{SINR}_{\text{min}}, \qquad\qquad\quad\:\,  \forall_{i, j}
\end{split}\\
\begin{split} \label{requestsconstraint}
{\sum_{j=1}^{{V}}} c_{i,j} \leq 1, \qquad\qquad\qquad\quad\qquad\quad\:\:\:\: \forall_{i}
\end{split}\\
\begin{split}\label{backhaulratecons}
{\sum_{i=1}^{{U}}}\sum_{j=1}^{{V}} r_{i,j}\cdot c_{i,j} \leq {B}_{R},
\end{split}
\end{align}
\end{subequations}
where the parameter $r_{i,j}$ is the requested data rate of $i^{th}$ RSU from $j^{th}$ child-drone. Also, $c_{i,j}$ is the optimization parameter, which takes the value 1 in case the link between $i^{th}$ RSU with $j^{th}$ child-drone is connected, and zero otherwise. Next, the objective function represents the sum-rate, that is, $S_R={\sum\limits_{i=1}^{U}}\sum\limits_{j=1}^{V} {r}_{i,j}$. The objective function is subject to the following constraints.

\begin{enumerate}
    \item Constraint \eqref{bwconstraint} shows that the $j^{th}$ child-drone can distribute the maximum bandwidth of $W_{j}$ to the RSUs.
    \item Constraint \eqref{linkconstraint} represents that a $j^{th}$ child-drone can serve up to $\tau_{j}$ RSUs.
    \item Constraint \eqref{powerconstraint} limits the optimal transmit power, $p_{i,j}$, of a $j^{th}$ child-drone for energy efficiency. Besides, constraint \eqref{Intconstraint} is the interference threshold for the $i^{th}$ RSU, where $\zeta_{i,j}$ is the product of the magnitude squared of the channel gain and the inverse of the path loss between the $i^{th}$ RSU and $j^{th}$ child-drone.
    \item Next constraint, i.e., \eqref{SINRconstraint} ensures that the communication link between the child-drone and the RSU satisfies the QoS requirement and is considered using the minimum SINR criterion.
    \item Constraint \eqref{requestsconstraint} ensures that an RSU is associated with only one child-drone.
    \item Constraint \eqref{backhaulratecons} implies that the achieved sum-rate of associated RSUs is within the backhaul data rate limit, $B_{R}$, which is the communication link between the parent-drone and the ground core-network. 
\end{enumerate}
\par It can be observed from the above constraints that, we considered interference as noise and maximum transmit power limit is also applied on a child-drone, which makes the optimization objective non-convex \cite{opt_1, opt_2, opt_3}. Besides, considering the objective function that maximizes the sum-rate of overall network, energy efficiency of the network in bits/s/Watt is defined by
\begin{equation}\label{ee}
    {EE}=\frac{S_{R}}{\eta\cdot{\sum\limits_{i=1}^{U}\sum\limits_{j=1}^{V} {p}_{i, j}\cdot c_{i, j}
 + \Lambda \times \Upsilon_{i,j}^{c} }},
\end{equation}
where $\eta$ is the inverse of power amplifier efficiency. Also, $\Lambda$ is the total number of associated RSUs, where $\Lambda\leq U$. Further, $\Upsilon_{i,j}^{c}$ is the circuit power cost by communication link between the $i^{th}$ RSU and the $j^{th}$ child-drone.

\subsection{Association of RSUs with Child-drones}
Having the distribution of the RSUs and the child-drones, each RSU calculates the SINR by using \eqref{sinr}. Also, each child-drone calculates the required bandwidth of RSUs using \eqref{eqbw} based on their demanded data rates. Next, the calculated parameters, threshold values, demanded data rates of the RSUs, and the total number of the RSUs and child-drones are passed as an input to Algorithm \ref{algo}, where it performs following steps to associate the RSUs with the child-drones.

\begin{enumerate}
    \item Each RSU selects a single child-drone out of $U$ child-drones, which gives the highest SINR value (Lines 1-2). Therefore, mathematically, a matrix ${\mathbf{M}}$ contains a number of ones in each column, in which each entry represents the connectivity between the $i^{th}$ RSU and the $j^{th}$ child-drone.
    \item At this step (Lines 3-10), each child-drone initializes two counters, $T_{\tau}$ and $T_W$, to zero, which represent the total number of associated requests and bandwidth distribution, respectively. Next, each child-drone keeps on associating the RSUs with itself considering the availability of bandwidth resource,  $W_{j}$, and the number of requests,  $\tau_{j}$. Here, it is important to note that a child-drone picks the RSU with the highest spectral efficiency first in the association process, where the spectral efficiency is defined as
    \begin{equation}\label{} 
\log_{2}(1+\text{SINR})=\frac{r}{{{w}}}.
\end{equation}
    The reason for selecting the the RSU with the best spectral efficiency first is because the objective of our problem discussed is Section \ref{objectivenet} is to maximize the sum-rate of the overall network. Therefore, having a higher value of numerator (data rate) and lower value of the denominator in the above equation will result in achieving the objective function more quickly.
    \item Finally, at the last step, the gathered information of association by all the child-drones is forwarded to the parent-drone, where the parent-drone considers the backhaul data rate constraint, $B_{R}$. If the constraint is satisfied, the association process completes, otherwise the parent-drone re-performs the steps mentioned in Lines 11-14 of Algorithm\,\ref{algo}.
\end{enumerate}
The complete algorithm for the association of RSUs with child-drones is summarized in Algorithm \ref{algo}.
   
\begin{algorithm}[t!]
 \label{algo}
\DontPrintSemicolon
  \KwInput{$U$, $V$, $r_{i,j}$, $w_{i,j}$,$W$, $\tau$, $\text{SINR}_{i,j}$, ${B}_{R}$}
  \KwOutput{{$ {\mathbf{M}}$}}
  \KwInitialize{$ {\mathbf{M}}= \text{\o}$}
    {\tcp{\textcolor{black}{Step at each RSU for the selection of child-drone from which it receives highest SINR value}}}  
  \For{$i=1$ \textbf{to} ${U}$}
  {
 Select child-drone from which a $i^{th}$ RSU receives maximum $\text{SINR}_\text{{max}}$\;
  }
    {\tcp{\textcolor{black}{Step at each child-drone for the association of RSUs with itself}}}  
    \For{$j=1$ \textbf{to} ${V}$}
  {
    \textbf{Initialize counters:} $ {T_{\tau}}=0$, $ {T_{W}}=0$\;
$j^{th}$ child-drone selects highest spectral efficient RSU from its list\;
  \leavevmode\newline
     \While{ $ {T_{\tau}} < \tau \wedge  {T_{W}} <  W$}
   {
           
             \If{$ {T_{W}} + {w_{i,j}} \leq  W$}
    {
        
        Update ${c_{i, j}} = 1$, $ {T_{\tau}}$ =$  {T_{\tau}} + 1$ and $ {T_{W}}$ = $  {T_{W}}$ + ${w_{i,j}}$\;
                        \Else
    {
{\textbf{break}}\;
    }        
        }
   }
  }
  {\tcp{\textcolor{black}{Step at the parent-drone to consider the constraint \eqref{backhaulratecons}}}}  
   \KwInitialize{$ {S_{R}}$ as total sum-rate of associated RSUs} 
    \While{$ {S_{R}}$ $>$ ${B}_{R}$}
   {
   \leavevmode\newline
   Select child-drone with max. associated RSUs\;{\tcp{\textcolor{black}{This approach of selecting child-drone will introduce fairness to entire region}}}
   Select the RSU with minimum data rate request for de-association\;{\tcp{\textcolor{black}{Because we want to maximize the sum-rate}}}
De-associate the selected pair and update ${c_{i, j}} = 0$, \;
$ {T_{\tau}}$ = $ {T_{\tau}} - 1$, $ {S_{R}}$ = $ {S_{R}} -$ ${r_{i,j}}$ and $ {T_{W}}$ = $ {T_{W}} -$ ${w_{i,j}}$
   }
\caption{Association of RSUs with child-drones.}
\end{algorithm}

\section{Performance Evaluation}\label{performance}
 In this section, we first explain the simulation scenarios including system parameters and then present simulation results, through which the performance of the our proposed system is evaluated. 
 
 \subsection{Simulation Description}
\begin{figure}
\centering
\includegraphics[scale=0.58]{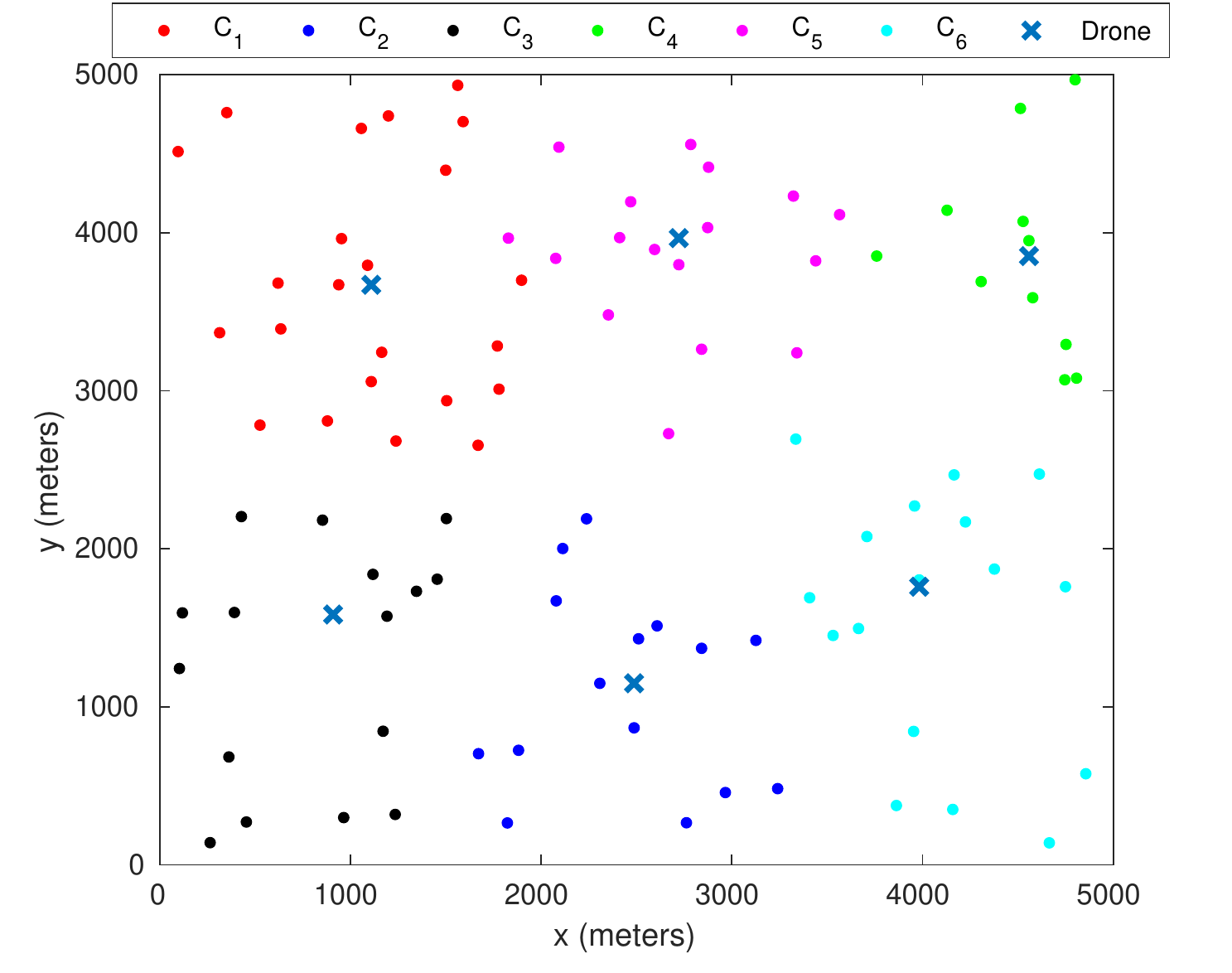}
\caption{2D view of distributions of the RSUs and positioning of the child-drones by \eqref{eq:8}.}
\label{fig1}
\end{figure}

Blockchain algorithm is tested and deployed on the personal Ethereum blockchain by setting the block gas limit up to 6 millions and following Petersburg hard fork. To feed the data to the smart contract, we use web3 APIs and Python 3. Moreover for the interaction between blockchain and our MATLAB simulation environment we used the Python 3 interface by creating TCP/IP ports between Python and MATLAB. Python 3 stores/retrieves the data to/from the blockchain (using web3), where MATLAB receives and sends data through these ports. In addition, an urban area of 25 km$^2$ is considered, where the RSUs are distributed according to $\textit{Matern type-I}$ hard-core process with the density of $\delta=  5\times10^{-6}$ per 1m$^2$, and the distance between the two RSUs is kept to $\zeta_{RSU}= 200$ m. Next, considering the location of the RSUs, $V=6$ drones are positioned following \eqref{eq:8}, at the altitude of $h_{D}$. A snapshot of this process is shown in Fig. \ref{fig1}, where a two-dimensional (2D) view of the RSUs and the drones can be seen. In the figure, a group of RSUs with the same colors belongs to the same drones that has the nearest mean value. For instance, the RSUs indicated by the red circles belong to cluster 1 (i.e., $C_{1}$), and the respective drone in that particular cluster is hovering at the altitude of $h_{D} = 200$ m.

With the obtained positions of the RSUs and the drones, the SINR is calculated using the simulation parameters mentioned in Table \ref{table:I}. In addition, random data rate demands are assigned to all the RSUs from a predefined data rate vector ${{\mathbf{r_\textrm{RSU}}}}$, which is given in Table \ref{table:I}. Therefore, having the demanded data rates and the SINR values of each RSU, the required bandwidth by each of the RSUs is calculated using \eqref{eqbw}. Finally, the obtained aforementioned values and specific constraints are given to the input of Algorithm \ref{algo}, where the association of the RSUs with the drones is performed to maximize the objective function given in \eqref{objective}.

 \subsection{Simulation Results and Analysis}
\begin{table} {}
\caption{Simulation Parameters for DESVN.}
\centering
 \begin{tabular}{|c| c| c| c|}
 \hline
 \textbf{Parameters} & \textbf{Numerical Values} & \textbf{Parameters} & \textbf{Numerical Values} \\ 
 \hline
 $h_{D}$ & 200\,m & $B_{R}$ &  1.40\,Gbps \\ [1ex]
 \hline
 $\lambda_{c}$ & 0.15\,m & $\psi_{\textrm{min}}$ & -10\,\text{dB} \\ [1ex]
 \hline
 $\alpha$ & 9.61 & $\sigma_{{n}}$ & -125\,\text{dB} \\
 \hline
 $\beta$ & 0.16 & $\varepsilon^{L}$ & 1\,\text{dB}  \\
 \hline
  $\varepsilon^{N}$ & 20\,\text{dB} & $\Upsilon^{c} $ & 0.1\,Watt \\ 
 \hline
 $\delta$ & $5\times10^{-6}/ m^2$  & $\zeta_{RSU}$ & 200\,m\\
 \hline
  $W_{j}$ & 400\,MHz & $\tau_{j}$ & 20\\
 \hline$
 P_{j}^{max}$ & 1.5\,Watt&$\eta$&0.20 \\
 \hline
$\mathbf{r_\textrm{RSU}}$ &
      \multicolumn{3}{c|}{\{{5, 10, 15, 20, 25}\}\,Mbps}  \\
 \hline
\end{tabular}
\label{table:I}
\end{table}
\begin{table*}[ht]
\centering
 \caption{Outcome and Authentication Time with different evaluation parameters and number of users.}
 \begin{tabular}{||c c| c c c||} 
 \hline
 Evaluation Parameter & Output&No. of Users (SVs)&Auth. Time (S) with 10 Reg. SVs&Auth. Time (S) with 20 Reg. SVs\\ 
 \hline\hline
 Served RSUs (\%age) & 71.23& 10&$0.161\times10^{-3}$&$0.331\times10^{-3}$  \\ 
 Achieved $S_{R}$ (bps) & $1.39\times10^{9}$&20&$0.321\times10^{-3}$&$0.639\times10^{-3}$ \\
 Avg. $W$ Consumption (Hz) &$237.13\times10^{6}$ &30&$0.484\times10^{-3}$&$0.964\times10^{-3}$  \\
 EE (bps/Watt) &$99.2\times10^{6}$ &40&$0.647\times10^{-3}$&$1.3\times10^{-3}$\\
 -&-&50&$0.811\times10^{-3}$&$1.7\times10^{-3}$\\
 \hline
 \end{tabular}
 \label{table:II}
\end{table*}

Table \ref{table:II} shows simulation results in terms of various key performance metrics such as percentage of associated RSUs, achieved sum-rate, average bandwidth consumed by the child-drones, and achieved energy efficiency of the overall network. It is important to note that these results are obtained with the backhaul data rate limit of 1.40\,Gbps. As a result, 71.23\% of the RSUs are associated with the drones. Also, the sum-rate, which is the objective function, is maximized, while the average bandwidth consumed by drones is $237.13\times10^{6}$\,Hz. Further, the energy efficiency of the overall network is $99.2\times10^{6}$ bits/s/watt. We note that 28.77\% of the RSUs are not associated with drones, which is due to the backhaul data rate limit; thus, drones cannot serve more requests. We will further discuss the impact of the backhaul data rate, when we present the results in Fig.~\ref{fig3}.

Table \ref{table:II} also provides an insightful look to the authentication process and the processing time required to authenticate the requested users, which depends on the system speed. In the authentication process, the requested user provides its unique address, and if the address is matched with the list of the registered addresses, the user is verified and vice versa. In Table \ref{table:II}, we observe that as the number of the requested users increases, the time to authenticate these users also increases, because we have to scan through the whole array of the registered users which we get from the blockchain. 
A similar trend can also be seen for the authentication time of the drones and the RSUs, as these entities follow the same procedure for authentication.

\begin{figure}
\centering
\includegraphics[scale=0.59]{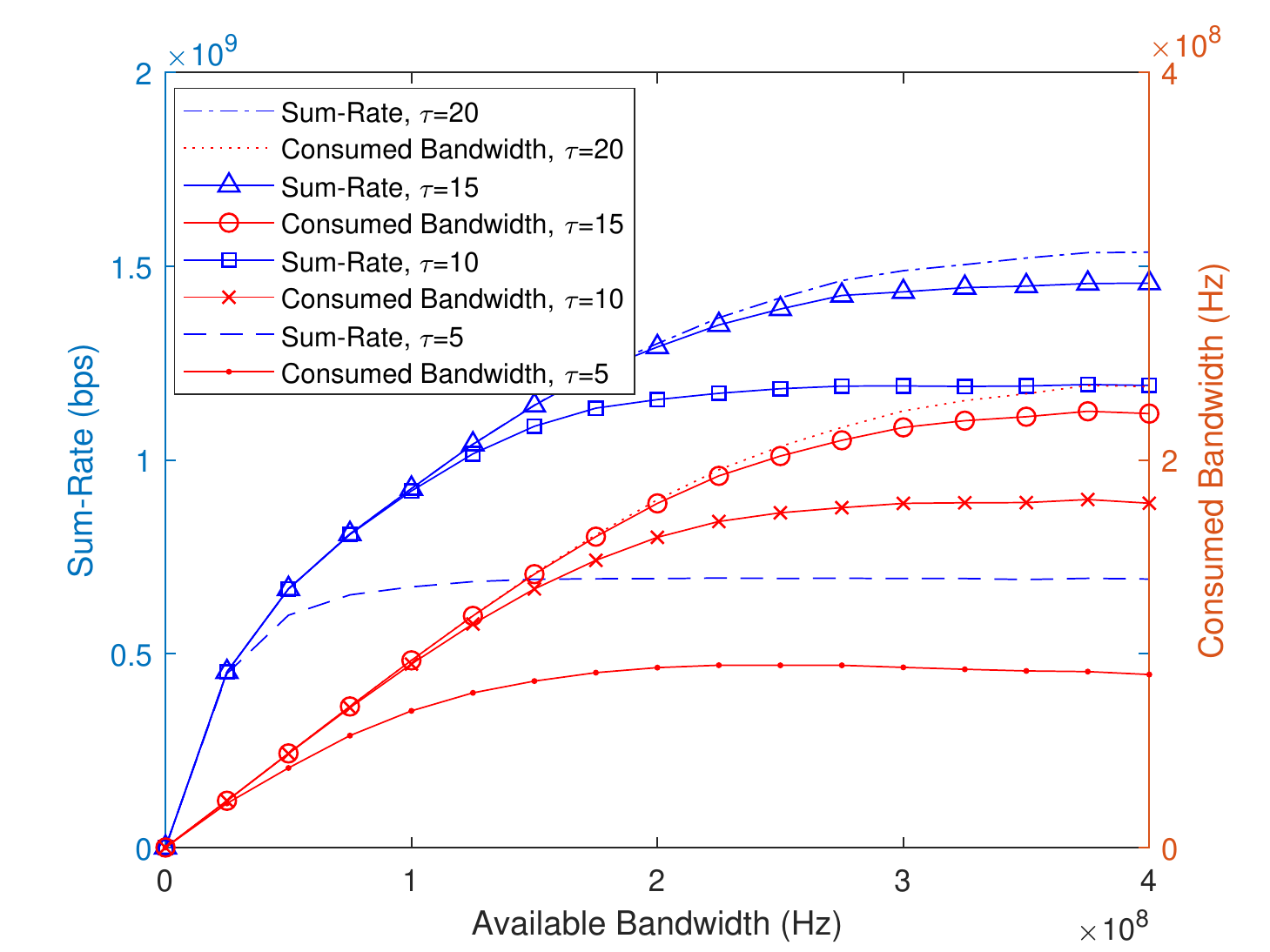}
\caption{Comparison of the consumed bandwidth and the sum-rate versus the available bandwidth $W_{j}$, when ${B_{R}}$= 1.60 Gbps and $\delta= 5\times10^{-6}/$m$^2$.}
\label{fig2}
\end{figure}

\begin{figure}
\centering
\includegraphics[scale=0.59]{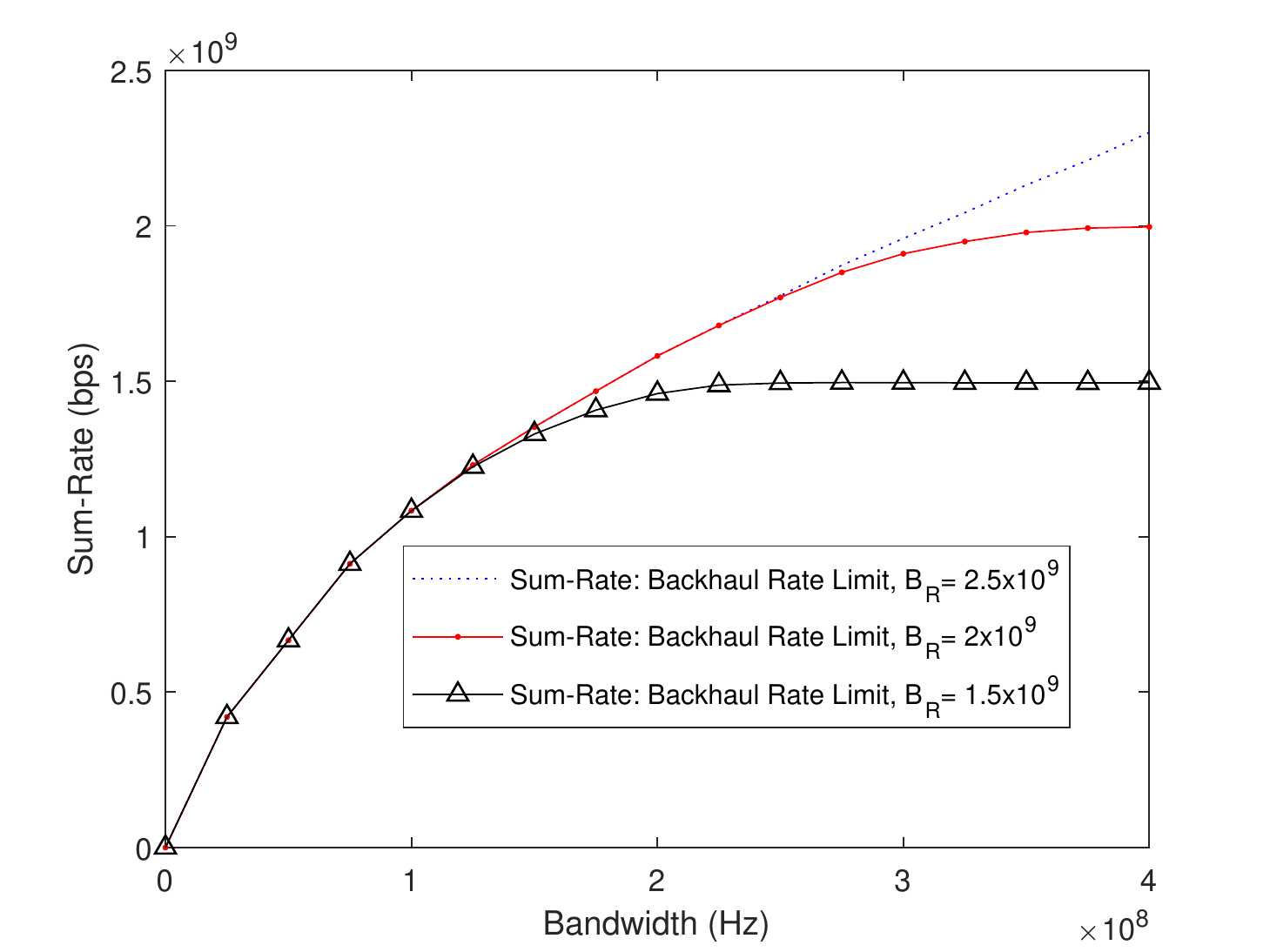}
\caption{Comparison of the sum-rate versus the available bandwidth $W_{j}$, when $\delta= 7\times10^{-6}$/m$^2$.}
\label{fig3}
\end{figure}

Fig. \ref{fig2} depicts the performance of the objective function (i.e., sum-rate), when the available bandwidth is increased from 0 to 400\,MHz. Also, a comparison of the consumed bandwidth versus the available bandwidth is shown in the figure. In the figure, the different curves correspond to the results with different numbers of the RSUs that a drone can support (i.e., $\tau=\{5, 10, 15, 20\}$). It can be observed that as the available bandwidth increases, the objective function increases. For example, when $\tau= 5$, the sum-rate increases from 0 to 0.67\,Gbps, and is constant onward, because a drone cannot serve more requests by the constraint of $\tau$. Similarly, the corresponding curve of the bandwidth consumption increases from 0 to 0.92 MHz, and then drones do not consume more bandwidth, since they cannot accommodate more RSUs. However, on the other side, increasing the value of $\tau$ results in higher sum-rate and bandwidth consumption, as expected.

Fig. \ref{fig3} shows the comparison of the achieved sum-rate versus the available bandwidth with different backhaul rate constraints $B_R$. It can be seen that the sum-rate increases with the increase in the bandwidth. For example, when the backhaul constraint is limited to 1.5\,Gbps, the sum-rate increases from 0 to 1.5\,Gbps, as the bandwidth increases. However, for the bandwidth larger than 200 MHz, the sum-rate saturates, because of the backhaul constraint $B_R$. However, when the backhaul constraint is increased to 2\,Gbps then sum-rate is also increased to 2\,Gbps. Similarly, the sum-rate further increases with $B_R=$ 2.5\,Gbps. It is important to note that, in case of 2.5\,Gbps backhaul constraint, the sum-rate is smaller than 2.5\,Gbps even with the bandwidth of 400\,MHz, which can be explained by the fact that the drones cannot serve more requests, because of its constraint $\tau$.

\begin{figure}
\centering
\includegraphics[scale=0.59]{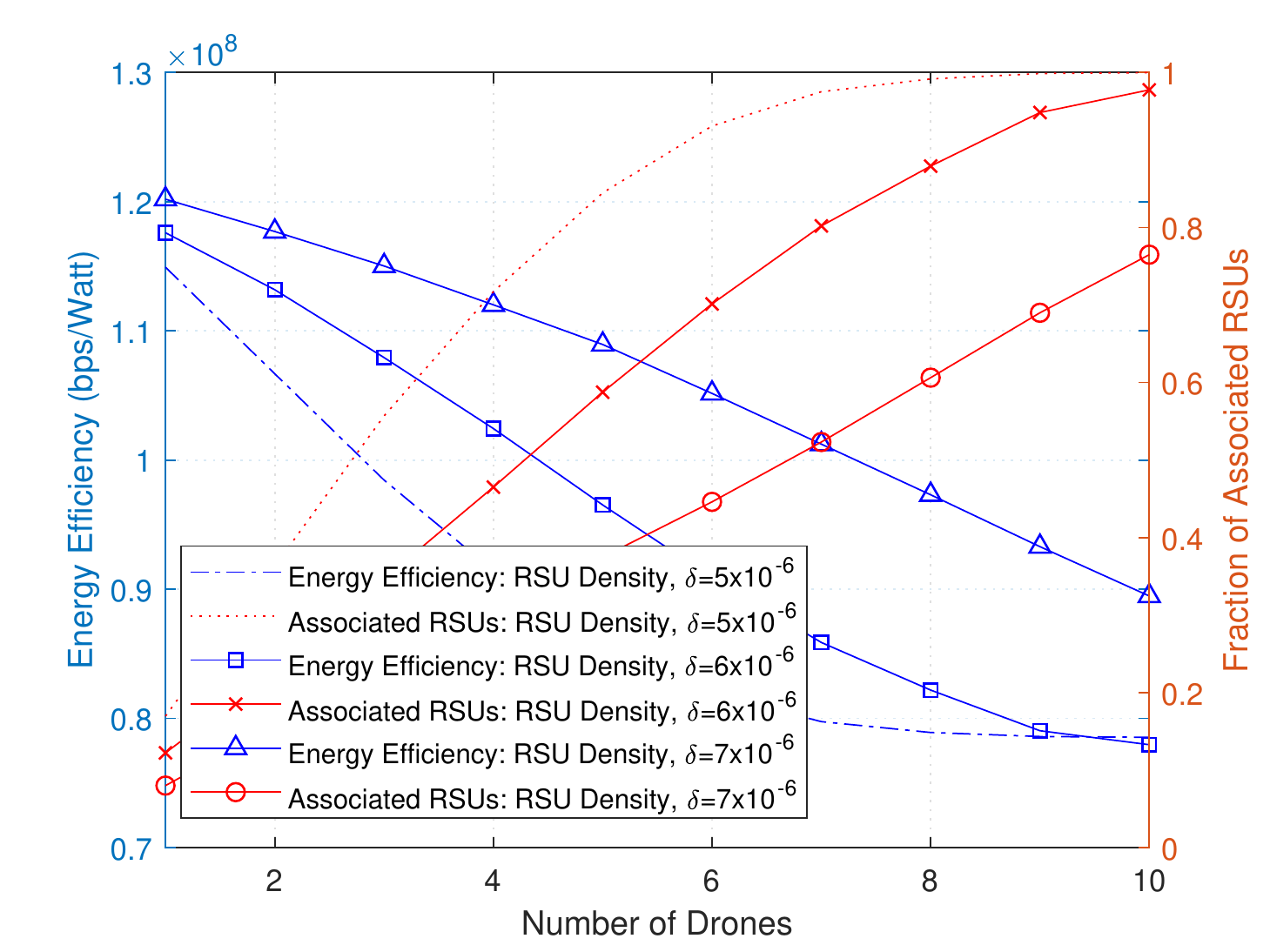}
\caption{Impacts of the number of the child-drones on the EE and the fraction of the associated RSUs with infinite $B_{R}$.}
\label{fig4}
\end{figure}

Fig. \ref{fig4} reveals the energy efficiency and the fraction of associated RSUs, when the number of child-drones is varied, while infinite $B_R$ is assumed. In the figures, the different curves corresponds to the different values of the RSUs density $\delta$. It can be observed that for each value of $\delta$, the energy efficiency decreases, while the the association ratio increases, when the number of drones is increased from 1 to 10. The reason for the increase in the fraction of the associated RSUs is that the coverage area increases, the number of drones increases. On the other side, the demanded values of the data rate of the RSUs are small, which makes the numerator of \eqref{ee} decline. Also, with higher number of the serving RSUs, the network consumes higher power, which means the greater denominator of \eqref{ee}. As a result, the energy efficiency decreases with the increase in the number of drones. Nevertheless, with the increase of the RSU density, higher data rate can be achieved, which gives higher energy efficiency. 

\begin{figure}
\centering
\includegraphics[scale=0.59]{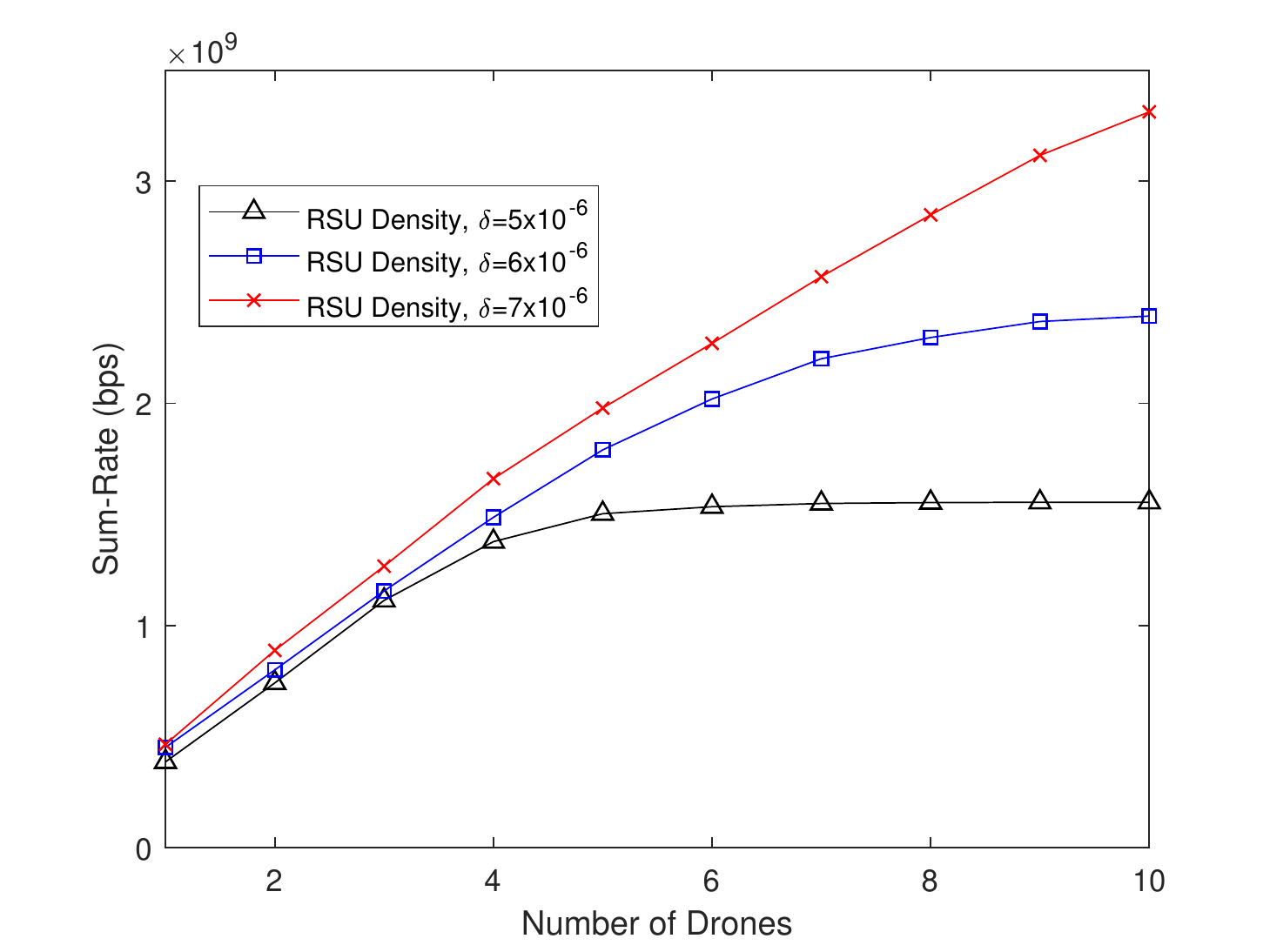}
\caption{Impact of the number of child-drones on the sum-rate with infinite $B_{R}$.}
\label{fig5}
\end{figure}

Fig. \ref{fig5} shows the impact of the number of drones being used on the achieved sum-rate changes for infinite $B_R$. In the figure, the three different curves indicate the results with different $\delta$, as in Fig.  \ref{fig4}. When $\delta=  5\times10^{-6}$/m$^2$, we observe that the sum-rate increases with the increase in the number of drones. However, after a certain point, the sum-rate does not increase, because all the RSUs are served, and hence there are no RSUs left. 
Also, with higher $\delta$, the sum-rates increases, because more RSUs are present to associate.

\begin{figure}
\centering
\includegraphics[scale=0.59]{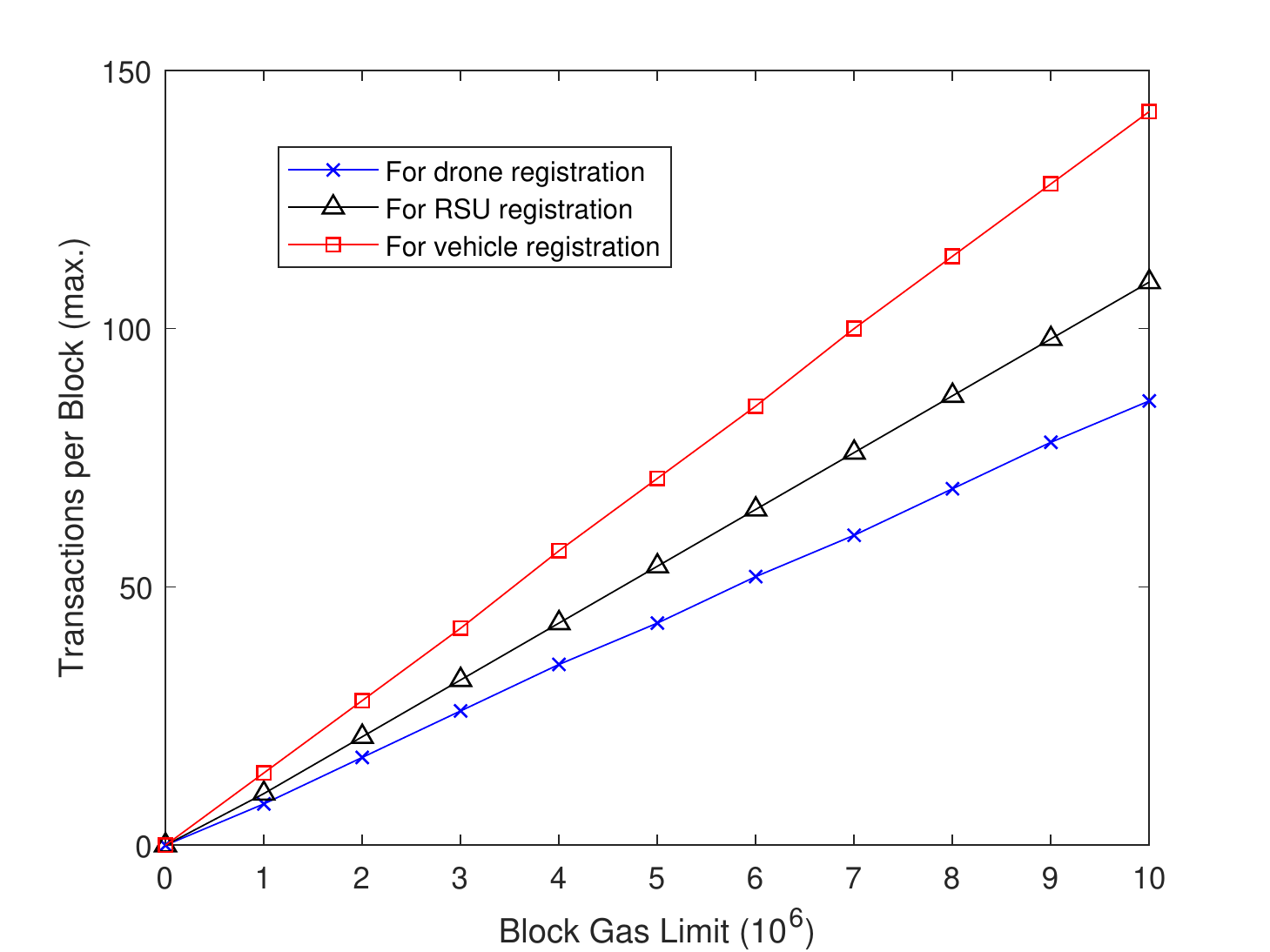}
\caption{Transactions per block versus block gas limit.}
\label{asaad}
\end{figure}

In Fig. \ref{asaad}, we investigate the registration process of the entities and provide the gas analysis of the transactions, which can only be made by the C\&C for the registration purposes. The gas used by each transaction function depends on the complexity of the function and the parameters required for that function. To get the gas required for the registration process of the drones, the smart vehicles, and the RSUs, we deploy the smart contract on the Ethereum blockchain and check the gas used by each transaction. It can be observed that only the C\&C is able to perform the registrations process (transactions made by the C\&C account register the entities). Fig. \ref{asaad} shows the number of transactions a block can accommodate during the registration process. We observe that as block gas limit increases, the number of transactions that can be accommodated by a single block also increases, because the number of transactions is directly related to the block gas limit. The same trend can be seen for the registration of each entity as each entity requires different parameters for the registration function. Therefore, we get gas analysis of each entity separately, and combined them in a single graph as shown in Fig. \ref{asaad}.

\begin{figure} 
\begin{center}
  \includegraphics[scale=0.43]{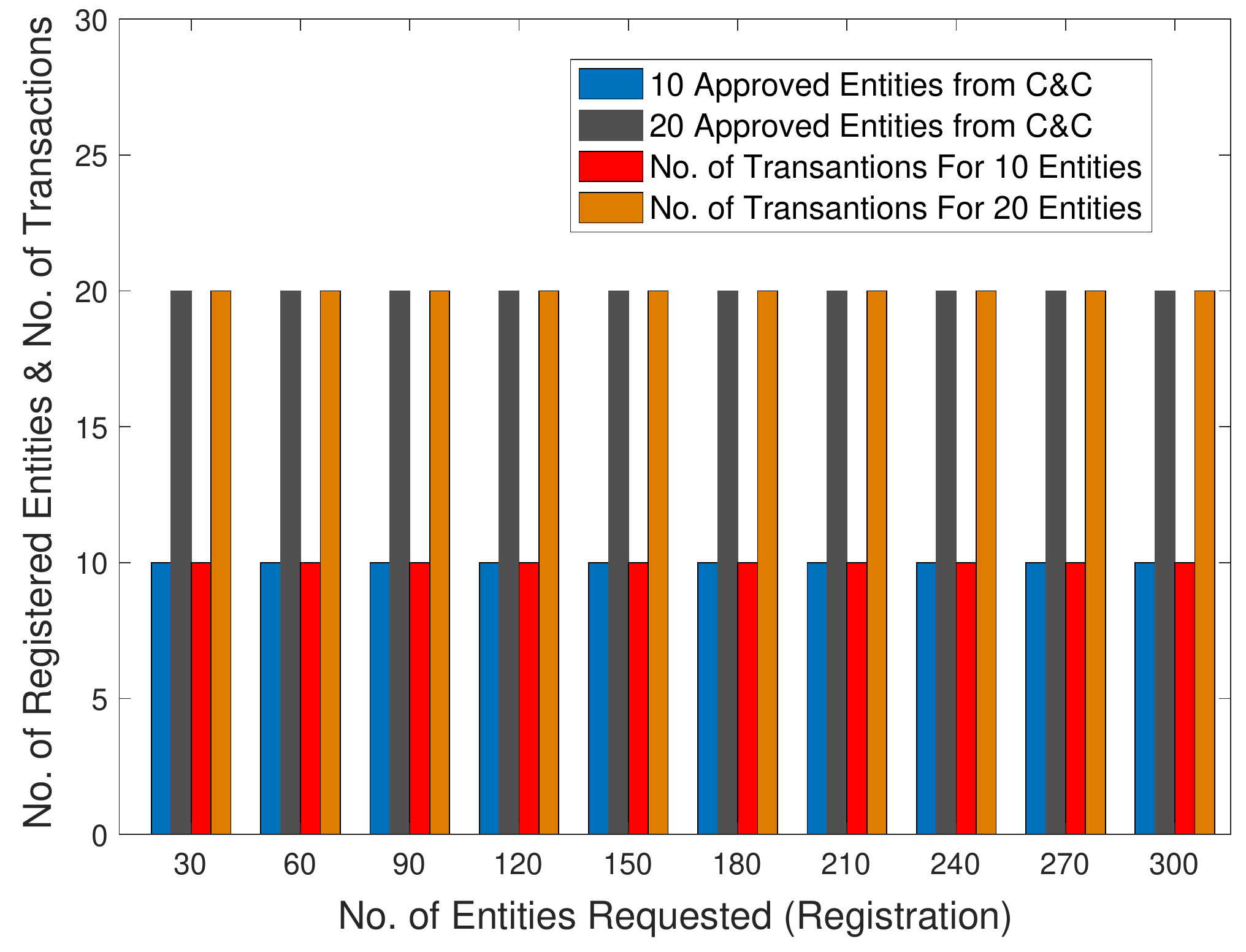}
  \caption{{Entities requested for registration versus registered entities.}}\label{flow}
\end{center}
\end{figure}

\begin{figure} 
\begin{center}
  \includegraphics[scale=0.46]{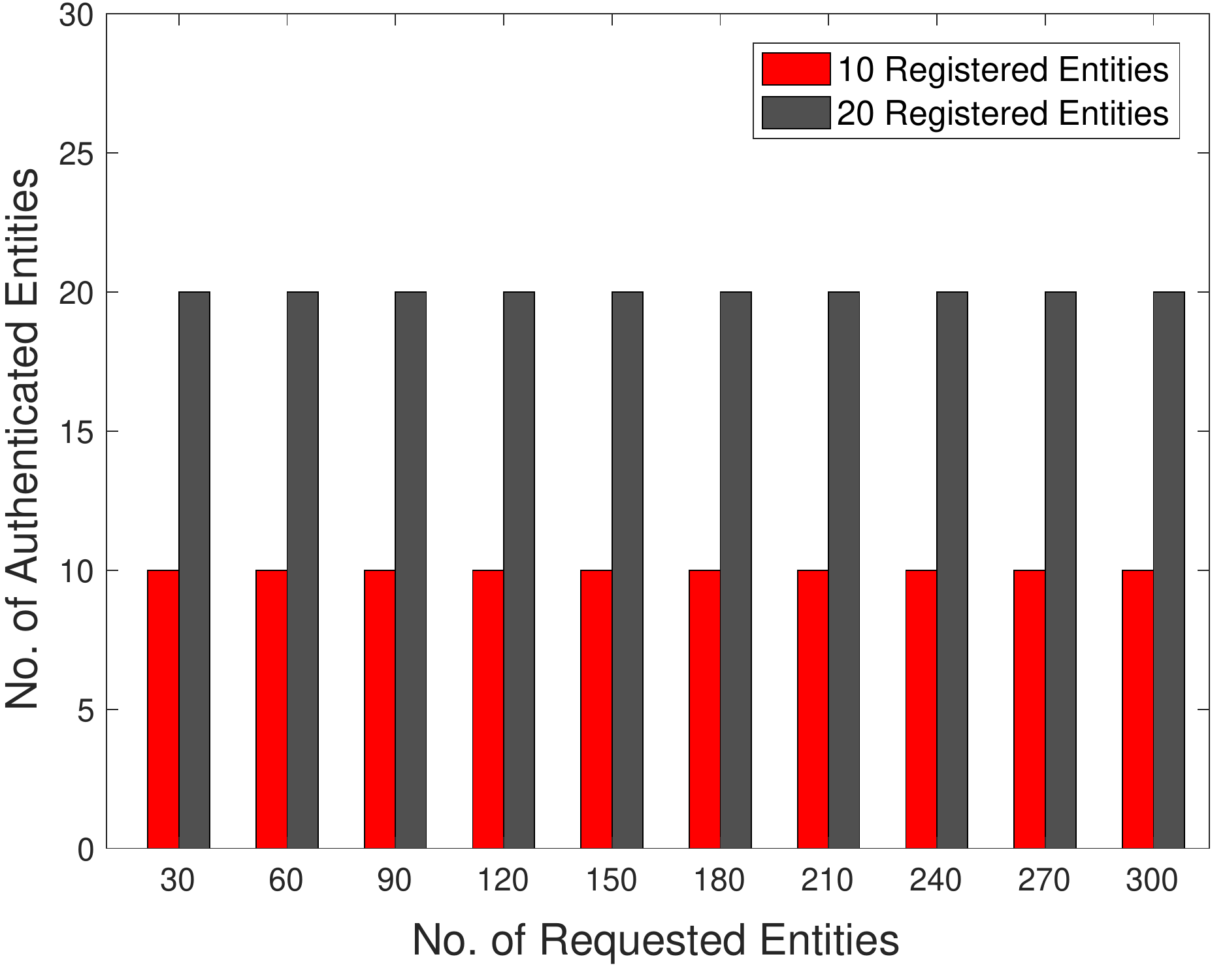}
  \caption{{Requested entities versus authenticated entities.} }\label{flow1}
\end{center}
\end{figure}

The same trend can also be seen for the registration and authentication process in Figs. \ref{flow} and \ref{flow1}. Fig. \ref{flow} shows that only the C\&C is able to register the entities no matter how many transactions are made by other accounts to register the entities. Moreover, it can also be seen that only those entities can get themselves registered, which are allowed by the C\&C irrespective of the number of requests of the entities for the registration purposes. On the other hand, only the registered entities can be authenticated by matching the information provided by the requested entity with the list of registered entities. In Fig. \ref{flow1}, regardless of the change in the number of requested, only the entities, which were registered by the C\&C at the first stage, can get the authentication.


\section{Conclusion and Future Directions} \label{conclusion}
In this paper, we have proposed a drone-aided secure vehicular network by employing blockchain technology and by placing the drones optimally in a geographic area. The blockchain is integrated to build mutual trust between different entities of the network and to protect the network from external intruders, who can be malicious. The algorithm also protects DESVNs by avoiding the single point of failure and centralized storage of the registered entities information, exploiting the distributed nature of the blockchain.

In the future, we are planning to analyze the limited storage capacity of the entities, as it has been observed that the blockchain size grows  with the increase in the number of the registered entities. Similarly, it can be investigated that how blockchain integration affects the overall energy of the system. Moreover, the concept of machine learning can be introduced to build intrusion detection systems and to tackle the unseen or zero day attacks as well the signature attacks in the system for further enhancement fo the system security. In addition, an effective firewall can be designed to filter the request, so that it is possible to take actions against the malicious requests from the adversaries once they are detected. Moreover, finding the optimal altitudes for different drones is worth investigating for an ever-increased spectral efficiency of the network.

\ifCLASSOPTIONcaptionsoff
  \newpage
\fi





\bibliographystyle{IEEEtran}

\vfill


\end{document}